# An Incremental Contact Model for Rough Surfaces of Strain Hardening Solids

**Yue Ding[#], Xuan-Ming Liang[#], Gang-Feng Wang***

Department of Engineering Mechanics, SVL and MMML, Xi'an Jiaotong University, Xi'an 710049, China

[#] These authors contribute equally to this work.

E-mail: wanggf@mail.xjtu.edu.cn

**Abstract**

The load-area relation of rough surfaces is of great interest in tribology. For elastic-plastic solids with strain hardening, an incremental model is adopted to analyze the contact of rough surfaces, in which the contact is modeled by accumulation of equivalent circular contacts with varying radius. For three typical rough surfaces with various material properties, comparisons with direct finite element calculations demonstrate the efficiency of this incremental contact model. An approximate linear relation between load and contact area is predicted by both methods up to a contact fraction of 15%. The influence of yield stress and strain hardening index on the load-area proportionality is presented. This work gives a simple while effective method to calculate the load-area relation for rough contact of strain hardening materials.

# 1. Introduction

The contact between rough solids is quite common in practice, for example, pistons against tubes [1], bolts against nuts [2], and tires against road surfaces [3]. Due to the existence of surface roughness, the real area in contact is often a small fraction of the projected area of the contact interface [4-6]. It has a close relationship with physical processes occur on the contact interface, such as friction, wear, lubrication, electrical conductivity and heat transfer [7-10]. Therefore, the determination of real contact area is one of the paramount issues in contact mechanics and tribology.

To make fair predictions of the rough contact process, extensive research has been conducted based on various assumptions and underlying mathematical frameworks, and the core of the approaches is the description of the surface morphology [11]. The elastic contact between rough surfaces was first modelled by Archard [12] in 1957. Regarding rough surfaces as protuberances superposed with smaller and smaller ones and applying spherical Hertzian solution for each of these protuberances, a power-law relation between normal load and the real contact area was predicted. Later, a multi-asperity contact model was developed by Greenwood and Williamson (GW) [13], in which rough surfaces are recognized as nominally flat surfaces covered with spherical asperities of equal radii and varying heights following a statistical distribution. Bush, Gibson and Thomas (BGT) [14] further treated the asperities as elliptical paraboloids with varying principal curvatures, and found that the normal load increases nearly proportional to the contact area. Using different sized spherical asperities to approximate such paraboloids, Greenwood [15] reproduced the BGT model in a simplified manner. For regular patterned rough surfaces, Li et al. [16] analyzed the contact stiffness considering the interactions among asperities. Other than the abovementioned multi-asperity theories, another approach was proposed by Persson [6]. The pressure distribution and subsequently the contact area under any normal load were derived from the power spectral density (PSD) of rough surface, regardless of the specific surface morphology. Interestingly, for elastic rough contact at low load, both multi-asperity contact models and Persson's theory predict linearity of the load-area relation [17].

For realistic surfaces, plastic deformation is usually inevitable during contact due to the existence of surface roughness. For fully plastic contact, Bowden and Tabor [18] assumed a constant flow pressure within the contact regions, which leads to the proportionality between applied load and contact area. Based on the volume conservation [19, 20], physical quantity continuity [21, 22] and direct finite element simulation [23, 24] of the elastic-plastic deformation of single asperity, extensions of the seminal GW model have been conducted for the elastic-plastic deformation of rough contact surfaces. Uniquely, Majumdar and Bhushan [25] developed a contact model regarding rough profile as the superposition of cosinoidal waves, in which the relation between load and real contact area depends on both fractal dimension and characteristic length of the rough profiles. Besides, through introducing the yield stress as the upper limit of the contact pressure distribution, Persson [3] studied the rough contact problem considering both elastic and plastic deformation. Megalingam and Mayuram [26] carried out the finite element simulation of rough contact and compared the results with statistical contact model. However, the effect of strain hardening behavior was not considered in the above studies.

For isotropic bilinear hardening solids, Sahoo and Ghosh [27] studied the effect of roughness on the relation between real contact area and normal load through finite element methods. Gao et al. [28] simulated the contact between a rigid sphere and a rough solid with strain hardening for both loading and unloading processes. Through finite element calculation, Pei et al. [29] addressed the influence of plastic properties on the elastic-plastic contact between a rigid plane and a self-affine fractal surface, in which the mean pressure is evidently enlarged as the strain hardening exponent increases. Considering strain hardening effect and interaction among asperities, Belhadjamor et al. [30] examined the influence of skewness and kurtosis on the rough contact for non-Gaussian random surfaces. Using the truncation method, Zhai et al. [31] investigated the relation between contact stiffness and surface roughness. By adopting the mechanism-based strain gradient plasticity theory, Song et al. [32] analyzed the contact between a rigid plane and an elastic-plastic solid with a rough surface. For the contact experiment between a rough polystyrene sphere and a flat glass [33], only if hardening effect was considered, could the experimental result be explained. Nonetheless, there are few theoretical models well-established for the rough contact of strain hardening solids.

Recently, Wang et al. [34, 35] proposed an incremental model for rough contact of both elastic and elastic-perfectly plastic materials, which gives accurate predictions of contact responses. In this work, we extend the incremental contact model to strain hardening solids. Section 2 briefly describes the incremental equivalent contact model for rough surfaces. The contact stiffness of the strain hardening substrate is studied by the indentation of a flat-ended cylindrical punch in Section 3. The finite element models of contact between rough surfaces and a rigid plane are presented in Section 4. In Section 5, the comparisons of load-area relations between our incremental model and finite element simulations are presented for various surface topographies and material properties. The effect of strain hardening on mean contact pressure is discussed in details.

## 2. The incremental equivalent contact model

Recently, an incremental equivalent model was advanced to deal with the contact of rough surfaces [32, 33], which shows good agreement with direct finite element simulation. In this model, the contact of rough surfaces is modeled by accumulation of equivalent circular contacts with its radius obtained by geometrical analysis. In this work, the incremental contact model is employed to deal with the elastic-plastic solids with strain hardening.

The contact between the strain hardening substrate with rough surface and a rigid plane is illustrated in Fig. 1(a). The strain hardening behavior of the substrate follows the power-law hardening model, which relates the stress $\sigma$ to strain $\varepsilon$ as

$$\sigma = \begin{cases} E\varepsilon, & \varepsilon \leq \varepsilon_0 \\ \sigma_y \left(\varepsilon / \varepsilon_0\right)^n, & \varepsilon > \varepsilon_0 \end{cases} \qquad (1)$$

where $E$ is the elastic modulus, $\sigma_y$ is the yield stress, $\varepsilon_0$ is the yield strain, and $n$ is the strain hardening exponent. When $n=0$, the material exhibits an elastic-perfectly plastic behavior. As for the case *n* approaching 1, the stress-strain relation acts more like linear elasticity. For common metals, the hardening exponent *n* varies from 0 to 0.5.

The distance between the rigid plane and the mean plane of the rough surface is donated as separation *z*, and the corresponding external load is *P*. At a separation *z*, the real contact area $A_c(z)$ and the number of contact patches $N(z)$ are achieved through truncating the rough surface by a virtual plane. The projection of the contact regions is schematically plotted in Fig. 1(b). The projected contact patches with various sizes and shapes can be further simplified by circular regions with an identical radius *R* as shown in Fig. 1(c). With a total area of $A_c(z)$, the equivalent radius is calculated as

$$R(z) = \left[\frac{A_c(z)}{\pi N(z)}\right]^{1/2} \tag{2}$$

With a decrease of surface separation d*z*, the increment of load d*P* can be obtained through the contact stiffness d*P*/d*z*. For strain hardening substrate and neglecting the interaction between different contact patches, the contact stiffness of *N* circular contacts of radius *R* can be expressed by [35]

$$\frac{dP}{dz} = g\left(\frac{P/N}{\pi R^2 \sigma_y}, n\right) 2NE^* R \tag{3}$$

where *g* is a dimensionless function which will be discussed later in Section 3, and $E^*$ is the combined elastic modulus determined by the Young's modulus *E* and Poisson's ratio *v* as $E^*=E/(1-v^2)$.

By combining Eqs. (2) and (3), we achieve

$$\frac{dP}{dz} = g\left(\frac{P}{A_c \sigma_y}, n\right) \frac{2E^*}{\sqrt{\pi}} (A_c N)^{1/2} \tag{4}$$

When the value of separation *z* approaches infinity, the initial load *P* equals to zero. The total load can be achieved through numerically solving the differential Eq. (4) by the fourth-order Runge-Kutta method. More details about the application of this explicit iteration methods can be seen in our previous paper [35].

**3. Contact stiffness of strain hardening solids**

To achieve the contact stiffness of a circular contact in Eq. (3), the indentation of a flat-ended cylinder on strain hardening substrate is simulated through commercial finite element software, ABAQUS. The finite element model is illustrated in Fig. 2, which is comprised of an elastic-plastic deformable substrate and a rigid flat-ended cylindrical indenter with a radius $R$. An external load $L$ is applied on the indenter and results in an indent depth $d$.

The radius of the indenter is 0.1mm. The size of the substrate is modeled to be $120R \times 120R$ to exclude the boundary effect. The Young's modulus $E$ of elastic-plastic substrate is 91GPa, Poisson's ratio $v$ is set as 0.3, and the modulus $E^*=1/(1-v^2)$ is calculated as 100GPa. For different values of yield stress ($\sigma_y$=100MPa, 400MPa, 700MPa and 1000MPa), various strain hardening exponents $n$=0, 0.1, 0.2, 0.3, 0.4 and 0.5 are considered in the simulations, which covers the mechanical properties of common metals.

Four-node bilinear axisymmetric quadrilateral reduced integration elements (CAX4R) are used to discretize the substrate, and the mesh is refined near the contact region. To reduce the stress concentration, the fillet is adopted at the edge of the indenter. The radius of the fillet is selected as $0.01R$ to eliminate the effect of fillet on the overall response. The $z$-axial displacement is restricted at the bottom of the bulk and on the boundary $r=0$, the symmetric boundary condition is applied. The contact between the indenter and the bulk is set to be frictionless. The accuracy of simulation results has been guaranteed through convergence tests.

Similar to the flat-ended cylindrical indentation on elastic-perfectly plastic material in previous work [35], the mechanical response of indentation on strain hardening substrate can be described by the following formula

$$\frac{L}{\pi R^2 \sigma_y} = f\left(\frac{2E^*\delta}{\pi \sigma_y R}, n\right) \tag{5}$$

Through fitting the simulation results plotted in Fig. 3(a), the function $f(x, n)$ is given by a unified form

$$f(x,n) = x\left(1 + 0.0806 x^{3.15}\right)^{(0.28n - 0.247)} \tag{6}$$

It can be seen that before the mean contact pressure reaches the yield strength, which corresponds to the regime $L/(\pi R^2 \sigma_y) < 1$, the relation between normalized load and normalized depth is linear and Eq. (5) reduces to Sneddon's solution [36]. Due to the accumulation of the plastic deformation, the normalized load deviates from the Sneddon's prediction as the indentation increases. With the occurrence of strain hardening, the curves approach the linear elastic relation. The larger the exponent is, the closer the curve is to the elastic case.

The contact stiffness of the flat-ended cylindrical indentation on a strain hardening material can be derived by differentiating Eq. (5) as

$$\frac{dL}{d\delta} = g\left(\frac{L}{\pi R^2 \sigma_y}, n\right) 2E^* R \tag{7}$$

For the analysis in the next section, the function $g(y, n)$ is simplified in an explicit form

as

$$g(y,n) = \left(1 + a_n y^{b_n}\right)^{c_n} \tag{8}$$

The dependences of $a_n$, $b_n$ and $c_n$ on the exponent $n$ are shown in Fig. 3(b) and fitted by

$$\begin{aligned} a_n &= 0.463 - 0.434 e^{-0.939n} \\ b_n &= 1.851n + 3.887 \\ c_n &= -2.403\left(1 + n^{0.741}\right)^{-6.01} \end{aligned} \tag{9}$$

## 4. Finite element model of contact between flat plane and rough surfaces

To verify the efficiency of our model, the contact between rough surfaces and a smooth plane is simulated through ABAQUS. As shown in Fig.4, the rough surface has a square projection with length $l$, and is discretized by 256×256 nodes with identical spacing. The height of the bulk is modeled to be $l/4$. Both ten-node quadratic tetrahedral (C3D10) elements and eight-node linear hexahedral (C3D8I) elements are adopted to discretize the substrate and the total number of the elements is 668630. The bottom of the substrate is fixed in all directions. The plane is modeled to be a rigid body and the contact property between the plane and the bulk is set as frictionless. The mechanical properties of elastic-plastic substrate are $E^*$=100GPa and $\nu$=0.3. Various values of yield stress ($\sigma_y$=100MPa, 400MPa, 700MPa and 1000MPa) and strain hardening exponents ($n$=0, 0.1, 0.2, 0.3, 0.4 and 0.5) are adopted in the simulations.

Three different types of rough surfaces are adopted in our simulations, and their morphologies are shown in Fig .5(a)-(c). Both surfaces A and B are isotropic and the heights of their nodes are artificially generated by an open-source code in MATLAB. Surface A is self-affine fractal on all length scales, while surface B remains fractal only in a high-frequency span, as shown in Fig. 5(d). The power spectra of both surfaces are axisymmetric and the Hurst exponents are 0.5 and 0.8 for surfaces A and B, respectively. Surface C is obtained from a ground metallic surface measured by a white light interferometer (NanoMap-1000WLI, AEP). It has an anisotropic characteristic and the PSDs perpendicular and parallel to the grinding direction are shown in Fig. 5(d). The lengths of surfaces A, B and C are 200 μm, 200 μm and 52.2 μm, and the root-mean-square roughness for each surface is 0.500 μm, 1.00 μm and 0.103 μm, respectively.

## 5. Results and discussions

According to our numerical method presented in Section. 2, the calculation of the load requires both the contact fraction $A_c/A_0$ and the number of contact patches $N$, which are illustrated in Fig. 6. With the decrease of separation, the contact fraction remains increasing while the number of contact patches fluctuates, due to the coalescence of the asperities at small separation. Both the proposed contact model in section 2 and finite element simulation in section 4 are carried out to calculate the load-area relations of these rough surfaces.

The variation of normalized load with respect to the contact fraction for surfaces A, B and C are shown in Figs. (7-9), respectively. Among these figures, the numerical results calculated from our model are displayed by lines, and the FEM results are plotted by scatters. It can be seen that our proposed model agrees well with the finite element simulations, which verifies the accuracy of our method. For three different rough surfaces, the relations between contact load and contact area hold an approximately linear trend for not only elastic materials (solid lines) and elastic-perfectly plastic materials ($n=0$, dashed lines), but strain hardening materials as well.

For materials with a specific hardening exponent $n$ and different values of yield stress, the load-area relations of three rough surfaces are shown in Figs. 7(a), 8(a) and 9(a). At a given external load, the contact area of strain hardening materials is larger than that of elastic materials because of the plastic deformation, but less than that of elastic-perfectly plastic materials due to the strain hardening effect. With the increase of yield stress, the contact fraction becomes smaller. In Figs. 7(b), 8(b) and 9(b), the relations between normalized load and contact fraction for materials with different hardening exponents are displayed. For all rough surfaces, the load corresponding to a certain contact area increases as the hardening exponent increases, which indicates that substrate with a larger hardening exponent has a higher capacity of bearing normal load.

Then the influence of strain hardening on mean pressure within the contact region is investigated for three rough surfaces. The relations between mean pressure normalized by yield stress and contact fraction for $\sigma_y/E^*=0.001$ and $\sigma_y/E^*=0.007$ are displayed in Fig. 10. For contact fraction larger than 1%, the mean contact pressure stays almost constant until 15% of the total contact area, which is consistent with the proportionality between load and area demonstrated in Figs. (7-9). As shown in Fig. 10, the difference of mean contact pressure among three surfaces for elastic-perfectly plastic materials ($n=0$) can be neglected compared with strain hardening materials. As the hardening exponent increases, the discrepancy of mean pressure among three surfaces becomes larger for both $\sigma_y/E^*=0.001$ and $\sigma_y/E^*=0.007$. However, for a given hardening exponent, the mean pressure of material with $\sigma_y/E^*=0.007$ is smaller than that of material with $\sigma_y/E^*=0.001$. These indicate that at the plastic limit (i.e. small hardening exponents), mean pressure is significantly affected by material properties and the influence of surface morphology becomes reasonably weak. With the increase of hardening exponent, the morphology of the surface becomes increasingly important. Furthermore, the distribution of mean pressure for $n$ varying from 0 to 0.5 is narrower for material with $\sigma_y/E^*=0.007$ when compared with the case of $\sigma_y/E^*=0.001$. At small contact fractions ($A_c/A_0<0.01$), statistical fluctuations can be observed since the number of contact patches is quite small and the area is highly related to the specific surface.

In Fig. 11, the variations of mean contact pressure with hardening exponent ranging from 0 to 0.5 are evaluated for five different rough surfaces. The yield stress is fixed as $\sigma_y=0.01E^*$, and the mean pressure at $A_c/A_0=0.05$ is demonstrated here. When $n=0$, the mean contact pressure approaches the elastic-perfectly plastic case, as discussed before [35]. For surfaces A, B and C, the value of mean pressure rises approximately linearly with the increase of the hardening exponent but with different slope. Similar trend can be observed for two fractal rough surfaces from the results in Ref. [29]. Such discrepancies of slopes among these surfaces can be attributed to the varying surface morphologies.

For varying values of yield stress, the mean contact pressure is plotted in Fig. 12(a) for three surfaces at a particular contact area ($A_c/A_0=0.05$). As the yield stress increases sufficiently high, the mean pressure of surfaces with identical morphology converges to the same value irrespective of the specific hardening exponents. It can be seen that the mean pressure strongly depends on surface morphology when elastic deformation dominates. As for small yield stress when plastic deformation is dominant, the contact pressure is determined by both the surface morphology and strain hardening effect. It is noteworthy that the influence of surface morphology on the exact mean pressure becomes prominent as the hardening exponent increases and gradually fades out as $n$ shrinks. For a certain range of material properties ($\sigma_y/E^*<0.01$), the dependence of mean pressure on yield stress follows a power-law relation as $P/(E^*A_c) \propto (\sigma_y/E^*)^k$ for all surfaces. Within this regime, surfaces with identical hardening exponent can be described with the same value of parameter $k$ despite their different topography. This indicates that the slope only relates to the hardening exponent. To investigate the effect of strain hardening, the variation of slope with respect to hardening exponent is displayed in Fig. 12(b). The value of $k$ decreases monotonically with the increase of the hardening index, and the relation between them can be described by a linear function as

$$k = -0.963n + 0.876 \tag{10}$$

When $n$ equals to 0, the relation can be applicable for elastic-perfectly plastic solids and the slope $k$ becomes 0.876, which coincides with the finite element results from Ref. [29].

## 4. Conclusions

This paper presents an incremental model to predict the relation between load and contact area for strain hardening materials with rough surfaces. For three typical rough surfaces of solids with various material properties, numerical results from our proposed model show great agreement with finite element simulations. The relations between load and contact area stay approximately linear for contact fraction up to 0.15. During the compression process, the mean contact pressure, which corresponds to the slope of the load-area relation, firstly fluctuates and then holds almost constant. For a given value of $\sigma_y/E^*$, as the hardening exponent increases, the mean pressure and subsequently the load also increase for any specific contact area, which indicates that a larger hardening exponent corresponds to a higher capacity of bearing normal load. As the ratio $\sigma_y/E^*$ approaches 0.1, the mean pressure is determined by the surface morphology. When the ratio $\sigma_y/E^*$ varies below 0.01, the mean pressure depends on both surface topography and strain hardening behavior. Within this regime, the relation between pressure and $\sigma_y/E^*$ can be described by a power-law function with an exponent only relating to strain hardening exponent. The presented results provide an effective and convenient method to predict the load-area relation for rough contact of strain hardening materials.

**Acknowledgements**

Supports from the National Natural Science Foundation of China (Grant No. 11525209) are acknowledged. Y. Ding acknowledges the support from the National Natural Science Foundation of China (Grant No. 12102322) and the China Postdoctoral Science Foundation (No.2018M64097).

**References**

1. Hu, Y., et al., *Numerical Simulation of Piston Ring in Mixed Lubrication—A*

*Nonaxisymmetrical Analysis.* Journal of Tribology, 1994. **116**(3): p. 470-478.
2. Fu, M., et al., *An Improved Shell-Fastener Model for Modelling C/Sic Composite Bolted Joints with Rough Surface Characteristics.* Composite Structures, 2020. **251**: p. 112516.
3. Persson, B.N.J., *Elastoplastic Contact between Randomly Rough Surfaces.* Physical Review Letters, 2001. **87**(11): p. 116101.
4. Jacobs, T.D.B. and A. Martini, *Measuring and Understanding Contact Area at the Nanoscale: A Review.* Applied Mechanics Reviews, 2017. **69**(6): p. 060802.
5. Ghaednia, H., et al., *A Review of Elastic-Plastic Contact Mechanics.* Applied Mechanics Reviews, 2017. **69**(6): p. 060804.
6. Persson, B.N.J., *Theory of Rubber Friction and Contact Mechanics.* The Journal of Chemical Physics, 2001. **115**(8): p. 3840-3861.
7. Vakis, A.I., et al., *Modeling and Simulation in Tribology across Scales: An Overview.* Tribology International, 2018. **125**: p. 169-199.
8. Müser, M.H., et al., *Meeting the Contact-Mechanics Challenge.* Tribology Letters, 2017. **65**(4): p. 118.
9. Persson, B.N.J., et al., *On the Nature of Surface Roughness with Application to Contact Mechanics, Sealing, Rubber Friction and Adhesion.* Journal of Physics-Condensed Matter, 2005. **17**(1): p. R1-R62.
10. Persson, B.N.J., *Contact Mechanics for Randomly Rough Surfaces.* Surface Science Reports, 2006. **61**(4): p. 201-227.
11. Almqvist, A., et al., *Interfacial Separation between Elastic Solids with Randomly Rough Surfaces: Comparison between Theory and Numerical Techniques.* Journal of the Mechanics and Physics of Solids, 2011. **59**(11): p. 2355-2369.
12. Archard, J.F., *Elastic Deformation and the Laws of Friction.* Proceedings of the Royal Society of London Series a-Mathematical and Physical Sciences, 1957. **243**(1233): p. 190-205.
13. Greenwood, J.A. and J.B. Williamson, *Contact of Nominally Flat Surfaces.* Proceedings of the Royal Society of London Series a-Mathematical and Physical Sciences, 1966. **295**(1442): p. 300-319.
14. Bush, A.W., R.D. Gibson, and T.R. Thomas, *The Elastic Contact of A Rough Surface.* Wear, 1975. **35**(1): p. 87-111.
15. Greenwood, J.A., *A Simplified Elliptic Model of Rough Surface Contact.* Wear, 2006. **261**(2): p. 191-200.
16. Li, S., et al., *Contact Stiffness of Regularly Patterned Multi-Asperity Interfaces.* Journal of the Mechanics and Physics of Solids, 2018. **111**: p. 277-289.
17. Carbone, G. and F. Bottiglione, *Asperity Contact Theories: Do They Predict Linearity Between Contact Area And Load?* Journal of the Mechanics and Physics of Solids, 2008. **56**(8): p. 2555-2572.
18. Bowden, F.P. and D. Tabor, *The Area of Contact between Stationary and between Moving Surfaces.* Proceedings of the Royal Society of London Series a-Mathematical and Physical Sciences, 1939. **169**(A938): p. 0391-0413.
19. Chang, W.R., I. Etsion, and D.B. Bogy, *An Elastic-Plastic Model for the Contact of Rough Surfaces.* Journal of Tribology, 1987. **109**(2): p. 257-263.
20. Sepehri, A. and K. Farhang, *Closed-Form Equations for Three Dimensional Elastic-Plastic Contact of Nominally Flat Rough Surfaces.* Journal of Tribology, 2009. **131**(4): p. 041402.

21. Zhao, Y., D.M. Maietta, and L. Chang, *An Asperity Microcontact Model Incorporating the Transition From Elastic Deformation to Fully Plastic Flow.* Journal of Tribology, 1999. **122**(1): p. 86-93.
22. Jeng, Y.-R. and S.-R. Peng, *Elastic-Plastic Contact Behavior Considering Asperity Interactions for Surfaces With Various Height Distributions.* Journal of Tribology, 2005. **128**(2): p. 245-251.
23. Kogut, L. and I. Etsion, *A Finite Element Based Elastic-Plastic Model for the Contact of Rough Surfaces.* Tribology Transactions, 2003. **46**(3): p. 383-390.
24. Jackson, R.L. and I. Green, *A Statistical Model of Elasto-Plastic Asperity Contact between Rough Surfaces.* Tribology International, 2006. **39**(9): p. 906-914.
25. Majumdar, A. and B. Bhushan, *Fractal Model of Elastic-Plastic Contact Between Rough Surfaces.* Journal of Tribology, 1991. **113**(1): p. 1-11.
26. Megalingam, A. and M.M. Mayuram, *Comparative Contact Analysis Study of Finite Element Method Based Deterministic, Simplified Multi-Asperity and Modified Statistical Contact Models.* Journal of Tribology, 2012. **134**(1).
27. Sahoo, P. and N. Ghosh, *Finite Element Contact Analysis of Fractal Surfaces.* Journal of Physics D: Applied Physics, 2007. **40**(14): p. 4245-4252.
28. Gao, C., H. Proudhon, and M. Liu, *Three-Dimensional Finite Element Analysis of Shallow Indentation of Rough Strain-Hardening Surface.* Friction, 2019. **7**(6): p. 587-602.
29. Pei, L., et al., *Finite Element Modeling of Elasto-Plastic Contact Between Rough Surfaces.* Journal of the Mechanics and Physics of Solids, 2005. **53**(11): p. 2385-2409.
30. Belhadjamor, M., et al., *Numerical Study of Normal Contact Stiffness: Non-Gaussian Roughness and Elastic–Plastic Behavior.* Proceedings of the Institution of Mechanical Engineers, Part J: Journal of Engineering Tribology, 2020. **234**(9): p. 1368-1380.
31. Zhai, C., D. Hanaor, and Y. Gan, *Contact Stiffness of Multiscale Surfaces by Truncation Analysis.* International Journal of Mechanical Sciences, 2017. **131-132**: p. 305-316.
32. Song, H., E. Van der Giessen, and X. Liu, *Strain Gradient Plasticity Analysis of Elasto-Plastic Contact Between Rough Surfaces.* Journal of the Mechanics and Physics of Solids, 2016. **96**: p. 18-28.
33. Weber, B., et al., *Molecular Probes Reveal Deviations from Amontons' Law in Multi-Asperity Frictional Contacts.* Nature Communications, 2018. **9**(1): p. 888.
34. Wang, G.F., X.M. Liang, and D. Yan, *An Incremental Equivalent Circular Contact Model for Rough Surfaces.* Journal of Tribology, 2021. **143**(8).
35. Liang, X.M., et al., *Elastic-Perfectly Plastic Contact of Rough Surfaces: An Incremental Equivalent Circular Model.* Journal of Tribology, 2021. **144**(5).
36. Sneddon, I.N., *The Relation Between Load and Penetration in the Axisymmetric Boussinesq Problem for a Punch of Arbitrary Profile.* International journal of engineering science, 1965. **3**(1): p. 47-57.

**Figure captions:**

Fig. 1. Schematic of the equivalent circular contact model.

Fig. 2. Finite element model of the flat-ended cylindrical indentation.

Fig. 3. (a) The dependence of normalized load on the dimensionless indent depth. (b) The relations between fitting parameters and the hardening exponent.

Fig. 4. Finite element model of a substrate with a rough surface compressed by a rigid plane.

Fig. 5. The morphologies of three rough surfaces: (a) surface A, (b) surface B and (c) surface C, and (d) their power spectral density functions.

Fig. 6. The dependence of (a) contact area and (b) the number of contact patches on separation.

Fig. 7. Variation of normalized load with respect to the contact fraction for surface A with (a) $n=0.1$ and different values of $\sigma_y$, and (b) $\sigma_y= 700$MPa and different values of $n$.

Fig. 8. Variation of normalized load with respect to the contact fraction for surface B with (a) $n=0.3$ and different values of $\sigma_y$, and (b) $\sigma_y= 400$MPa and different values of $n$.

Fig. 9. Variation of normalized load with respect to the contact fraction for surface C with (a) $n=0.5$ and different values of $\sigma_y$, and (b) $\sigma_y= 1000$MPa and different values of $n$.

Fig. 10. The dimensionless mean contact pressure versus contact fraction for solids with $\sigma_y/E^*=0.001$ and $\sigma_y/E^*=0.007$.

Fig. 11. The variation of normalized mean contact pressure with respect to the hardening exponent.

Fig. 12. (a) The variation of normalized mean contact pressure with respect to the normalized yield stress. (b) The relations between the slope and hardening exponent.

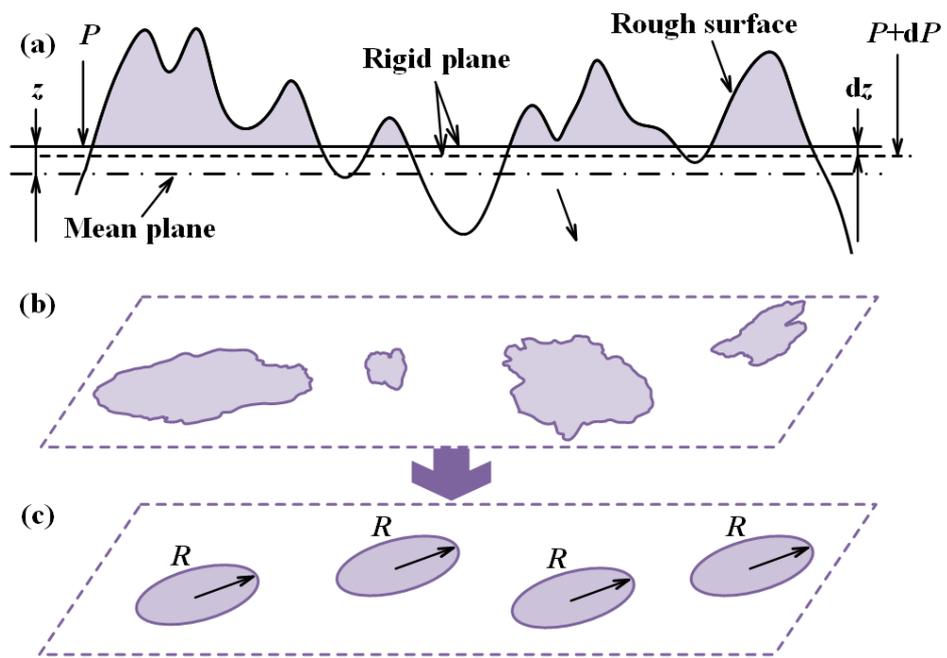

Fig. 1

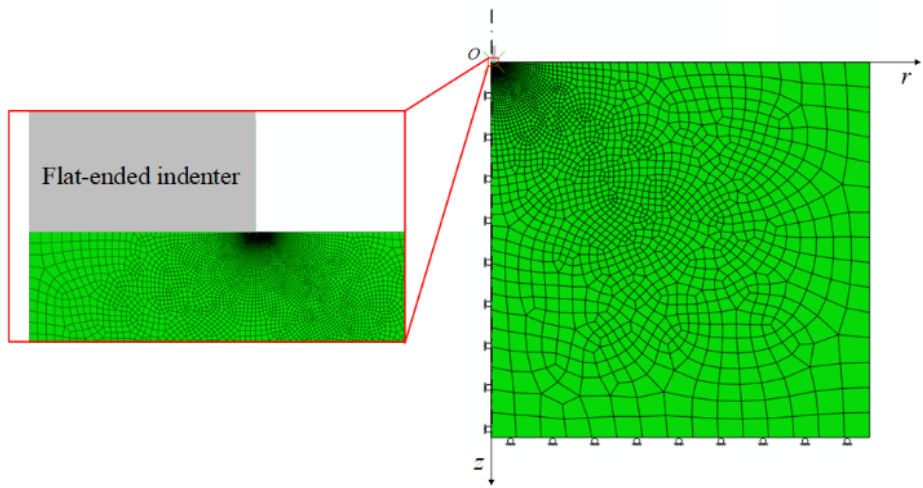

Fig. 2

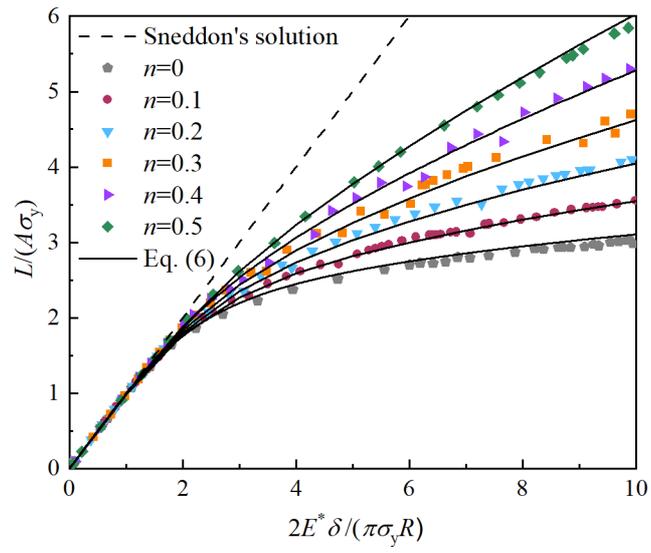

(a)

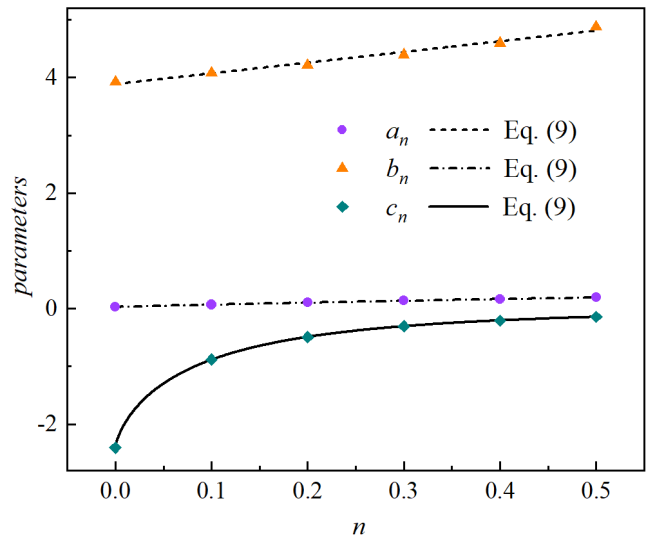

(b)

Fig. 3

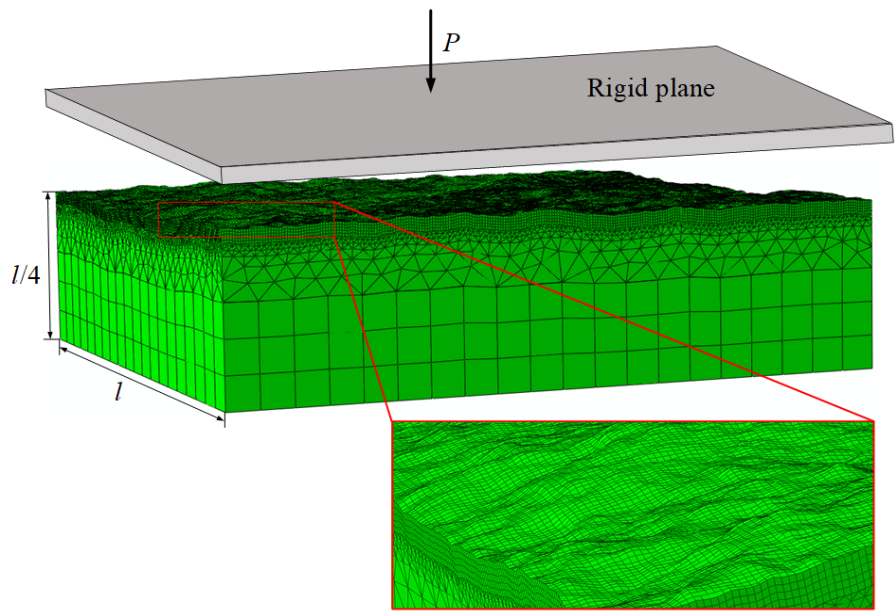

Fig. 4

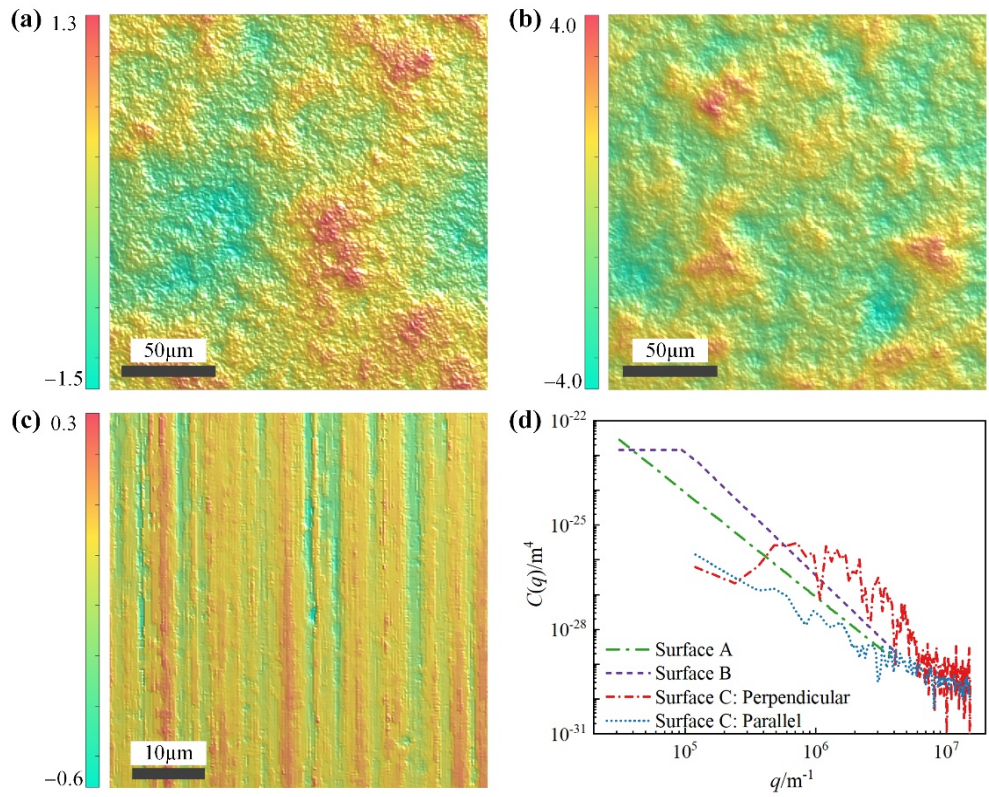

Fig. 5

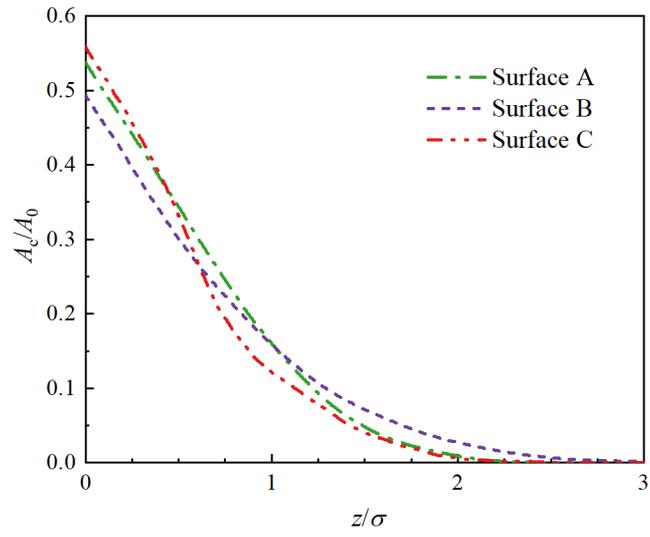

(a)

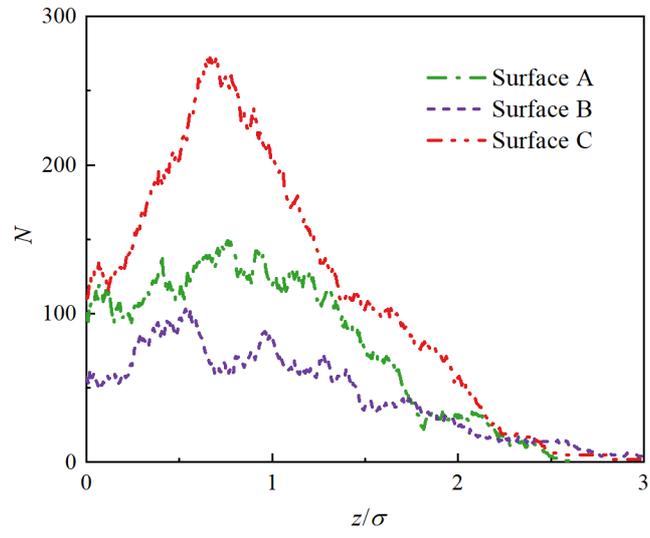

(b)

Fig. 6

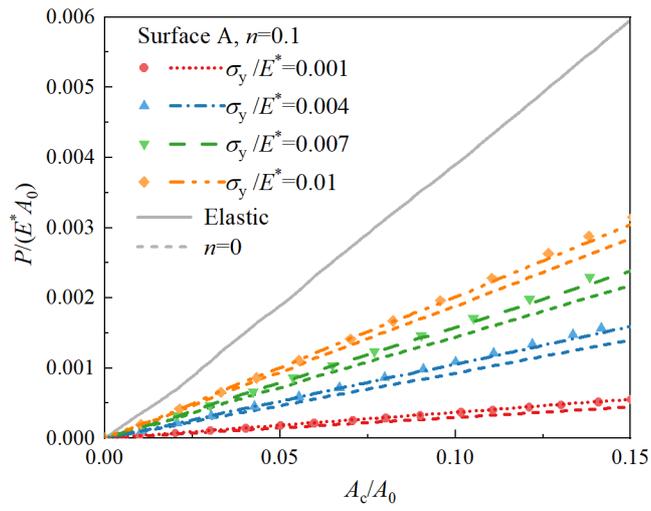

(a)

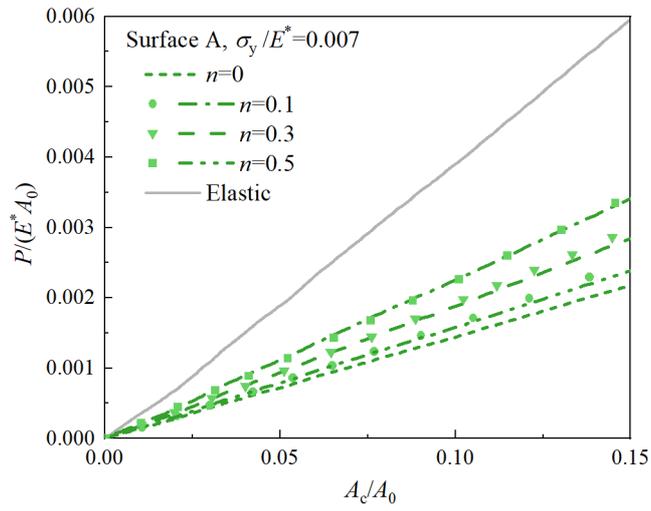

(b)

Fig. 7

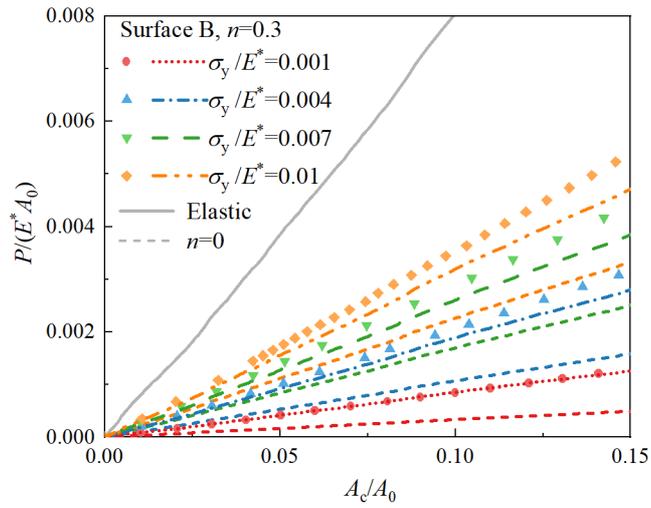

(a)

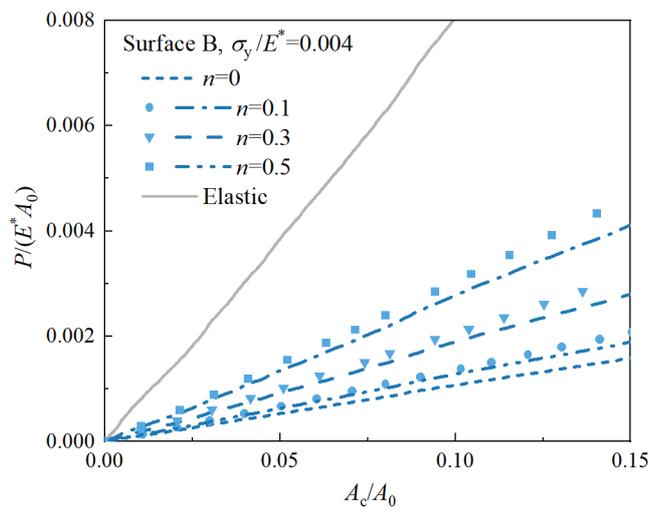

(b)

Fig. 8

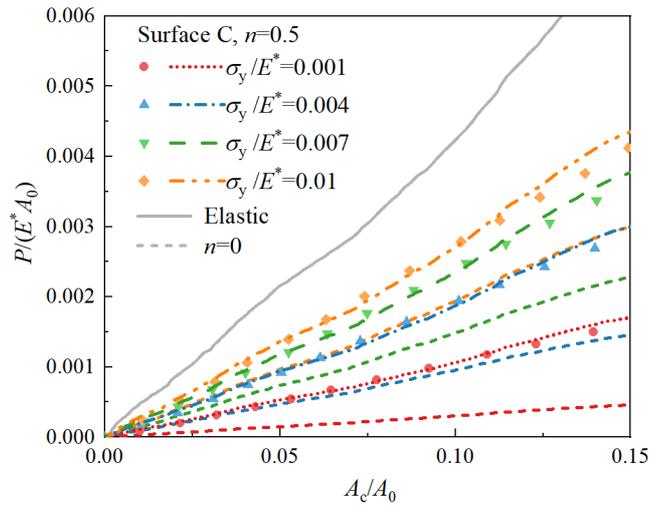

(a)

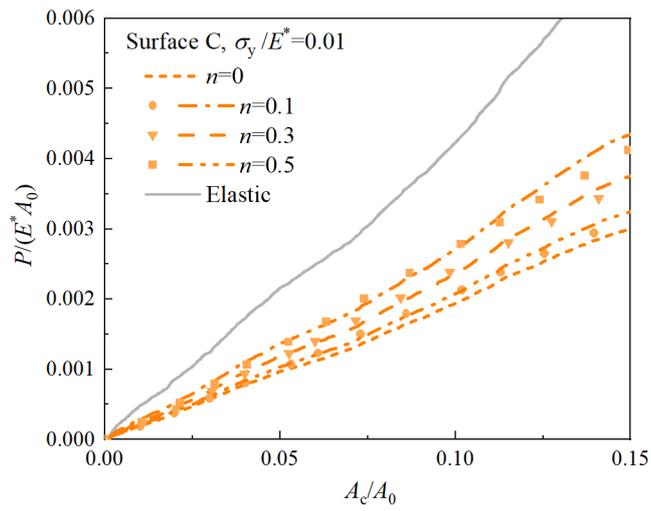

(b)

Fig. 9

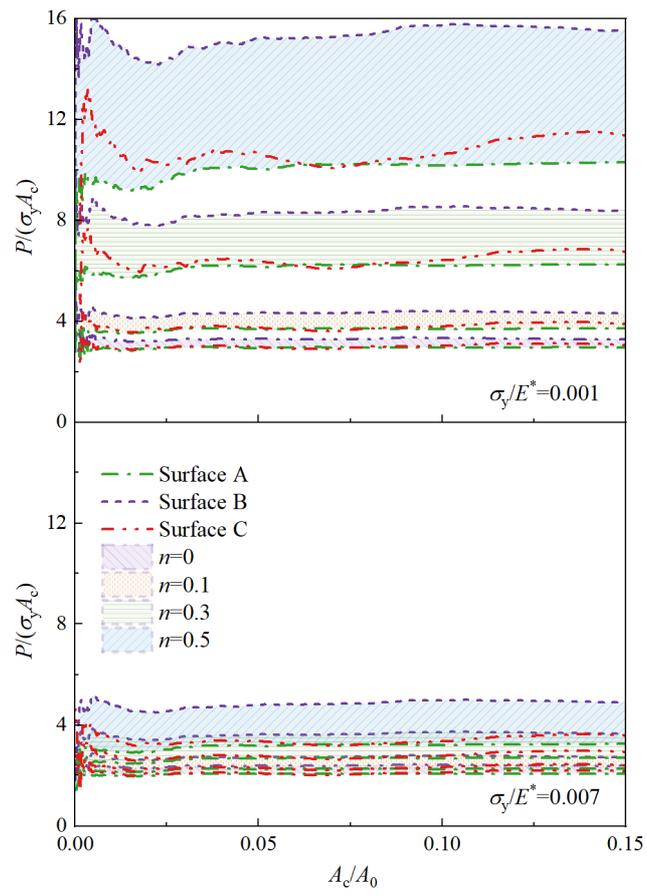

Fig. 10

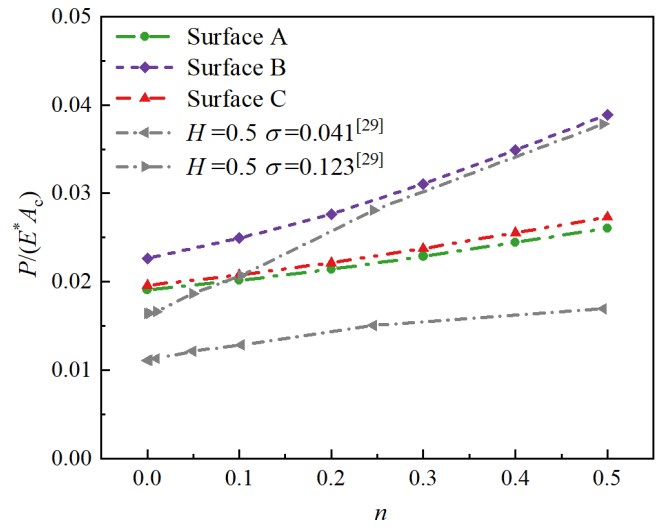

Fig. 11

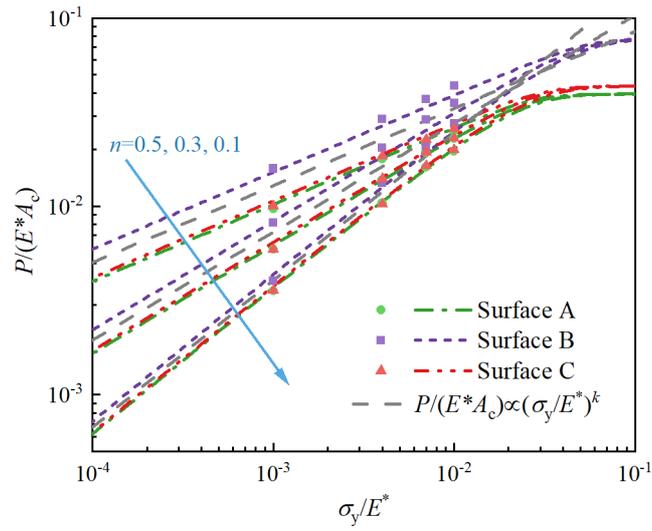

(a)

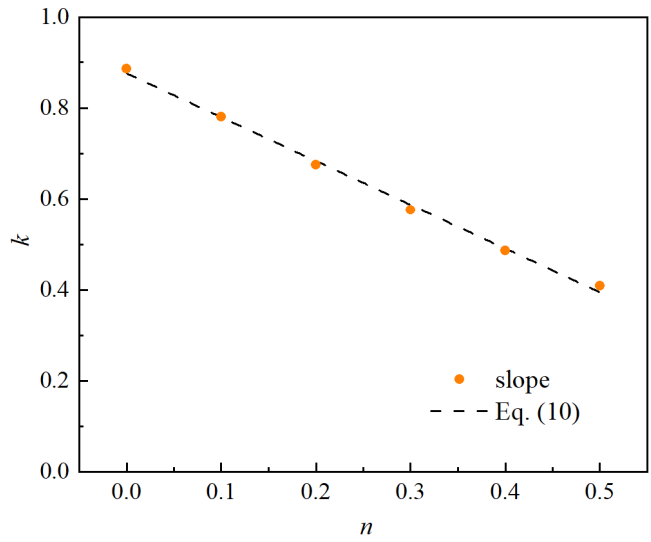

(b)

Fig. 12